\documentclass[longauth,letter]{aa}

\usepackage{graphicx}
\usepackage{txfonts}
\usepackage{amsmath,amssymb}
\usepackage{natbib}
\usepackage[squaren,Gray]{SIunits}
\usepackage{color, colortbl}
\usepackage{array}
\usepackage{hyperref}
\usepackage{longtable,lscape}
\usepackage{multirow}
\usepackage{hyperref}

\providecommand{\tabularnewline}{\\}

\newcommand{\Mjup}{$M_{\text{Jup}}$}

\newcommand\Lp{$\text{L}^\prime$}

\newcommand {\micron}{\unit{}{\micro\meter}}

\begin{document}

   \title{Discovery of a low-mass companion inside the debris ring surrounding the F5V star HD\,206893}

   \author{
         J. Milli \inst{1}
        \and    P. Hibon\inst{1}
         \and V. Christiaens\inst{2,3,4}
         \and \'E. Choquet\inst{5} 
         \and M. Bonnefoy\inst{6}
         \and G. M. Kennedy\inst{7}
         \and M. C. Wyatt \inst{7}
         \and O. Absil \inst{4} 
         \and C.~A.~G\'omez Gonz\'alez \inst{4}
         \and C. del Burgo \inst{8}
         \and L. Matr\`{a} \inst{7}
         \and J.-C. Augereau\inst{6}
         \and A. Boccaletti \inst{9}
         \and C. Delacroix \inst{10} 
         \and S. Ertel \inst{11}
        \and W. R. F. Dent \inst{12}
        \and P. Forsberg \inst{13}
        \and T. Fusco \inst{14}
        \and J. H. Girard \inst{1}
        \and S. Habraken \inst{4}
        \and E. Huby \inst{4}
        \and M. Karlsson \inst{13}
        \and A.-M. Lagrange\inst{6}
        \and D. Mawet\inst{5,16}
        \and D. Mouillet \inst{6}
        \and M. Perrin\inst{15}
        \and C. Pinte \inst{6}
        \and L. Pueyo\inst{15}
        \and C. Reyes\inst{1}
        \and R. Soummer\inst{15}
        \and J. Surdej \inst{4}
        \and Y. Tarricq \inst{1}
         \and Z. Wahhaj \inst{1}
          }

   \institute{European Southern Observatory (ESO), Alonso de C\'ordova 3107, Vitacura, Santiago, Chile, 
          \email{jmilli@eso.org}
        \and
            Departamento de Astronom\'ia, Universidad de Chile, Casilla 36-D, Santiago, Chile
        \and 
        Millenium Nucleus "Protoplanetary Disks in ALMA Early Science", Chile
        \and
        Space sciences, Technologies and Astrophysics Research (STAR) Institute, Universit\'e de Li\`ege, 19c All\'ee du Six Ao\^ut, B-4000 Li\`ege, Belgium.
         \and 
         Jet Propulsion Laboratory, California Institute of Technology, 4800 Oak Grove Drive, Pasadena, CA 91109, USA
         \and
            Univ. Grenoble Alpes, CNRS, IPAG, F-38000 Grenoble, France 
        \and     
            Institute of Astronomy, University of Cambridge, Madingley Road, Cambridge, UK 
            \and
            Instituto Nacional de Astrof\'{i}sica, \'{O}ptica y Electr\'{o}nica, Luis Enrique Erro 1, Sta. Ma. Tonantzintla, Puebla, Mexico
        \and
          LESIA, Observatoire de Paris, PSL Research University, CNRS, Sorbonne Universit\'es, UPMC Univ. Paris 06, Univ. Paris Diderot, Sorbonne Paris Cit\'e, 5 place Jules Janssen, 92195 Meudon, France
            \and Sibley School of Mechanical and Aerospace Engineering, Cornell University, Ithaca, USA
            \and Steward Observatory, University of Arizona, 933 N Cherry Ave, Tucson, AZ 85719, USA
            \and
            Atacama Large Millimeter/submillimeter Array (ALMA) Santiago Central Offices, Alonso de C\'{o}rdova 3107, Vitacura, Casilla 763 0355, Santiago, Chile
            \and 
            Department of Engineering Sciences, {\AA}ngstr\"{o}m Laboratory, Uppsala University, Box 534, 751 21 Uppsala, Sweden
        \and
          ONERA, The French Aerospace Lab, BP72, 29 avenue de la Division Leclerc, 92322 Chatillon Cedex, France
        \and 
          Space Telescope Science Institute, 3700 San Martin Drive, Baltimore, MD 21218 USA
         \and 
         Department of Astronomy, California Institute of Technology, 1200 E. California Blvd, MC 249-17, Pasadena, CA 91125 USA
        }

   \date{Received: 15 October 2016; accepted 28 November 2016 }

 
  \abstract
  {}
   {Uncovering the ingredients and the architecture of planetary systems is a very active field of research that has fuelled many new theories on giant planet formation, migration, composition, and interaction with the circumstellar environment. We aim at discovering and studying new such systems, to further expand our knowledge of how low-mass companions form and evolve.}
   {We obtained high-contrast H-band images of the circumstellar environment of the F5V star HD\,206893, known to host a debris disc  never detected in scattered light. These observations are part of the SPHERE High Angular Resolution Debris Disc Survey (SHARDDS) using the InfraRed Dual-band Imager and Spectrograph (IRDIS) installed on VLT/SPHERE.}
   {We report the detection of a source with a contrast of $3.6\times 10^{-5}$ in the H-band, orbiting at a projected separation of 270~milliarcsecond or 10\,au, corresponding to a mass in the range 24 to 73\Mjup{} for an age of the system in the range 0.2 to 2 Gyr. The detection was confirmed ten months later with VLT/NaCo, ruling out a background object with no proper motion. A faint extended emission compatible with the disc scattered light signal is also observed.}
   {The detection of a low-mass companion inside a massive debris disc makes this system an analog of other young planetary systems such as $\beta$ Pictoris, HR\,8799 or HD\,95086 and requires now further characterisation of both components to understand their interactions.}


 \maketitle

\section{Introduction}

Through direct imaging, instruments fed with adaptive optics (AO) have enabled the detection and characterisation of a few tens of low-mass companions, either giant planets (hereafter GP) or brown dwarfs (BD), probing a parameter space in the mass vs orbital radius still inaccessible with other indirect techniques such as radial velocities or transits. The direct detection of the thermal emission of such substellar objects brings precious information for understanding their formation mechanisms and physical properties \cite[see][for a recent review]{Bowler2016}. In addition, many of the GP/BD systems discovered in high-contrast imaging are associated to a debris disc, generally detected through its infrared or submillimetre emission (e.g. HR\,8799, \citealt{Marois2006};  HD\,95086, \citealt{Rameau2013_discovery}; HR\,3549 \citealt{Mawet2015}; HR~2562, \citealt{Konopacky2016}). In only three cases, this disc was also resolved in scattered light ($\beta$ Pictoris, HD\,106906 and Fomalhaut), enabling to study the interactions with the companion.

This letter presents the discovery of a low-mass BD in orbit around the nearby F5V star HD\,206893 located at $38.3\pm0.8$~pc (see details in Table \ref{tab_prop}). The star is known to host a debris disc detected through its large infrared excess \cite[$L_{dust}/L_\star=2.3 \times 10^{-4}$,][]{Moor2006}, characterised through its Spectral Energy Distribution (SED) with Spitzer/IRS-MIPS \citep{Chen2014}, and marginally resolved with Herschel/PACS (as detailed in this letter). This study also presents the putative scattered light signal of the disc, at low signal-to-noise (S/N) due to the faintness of its emission. The age of the system is not well constrained as it is not known to belong to a moving group. \citet{Zuckerman2004} estimated an age of 200 Myr based on X ray, radial velocity and proper motion measurements. This age is also inferred by \citet{Holmberg2009} with an upper limit of 1.2 Gyr from Padova stellar evolution models. More recently,  \citet{David2015} suggest a median age of 2.1 Gyr using a Str\"omgren photometry fit to stellar atmosphere models in a Bayesian framework while \citet{Pace2013} derives $860 \pm 710$ Myr based on the chromospheric activity calibrated against the Geneva-Copenhagen survey. 

\section{Observation and data reduction}
\label{sec_data_red}

The companion was first detected with VLT/SPHERE in 2015. HST/NICMOS archival data from 2007 showed that the companion cannot be a background object without proper motion and data from VLT/NaCo redetected the object in 2016.

\subsection{SPHERE}
\begin{figure}
        \centering
        \includegraphics[width=0.9\hsize]{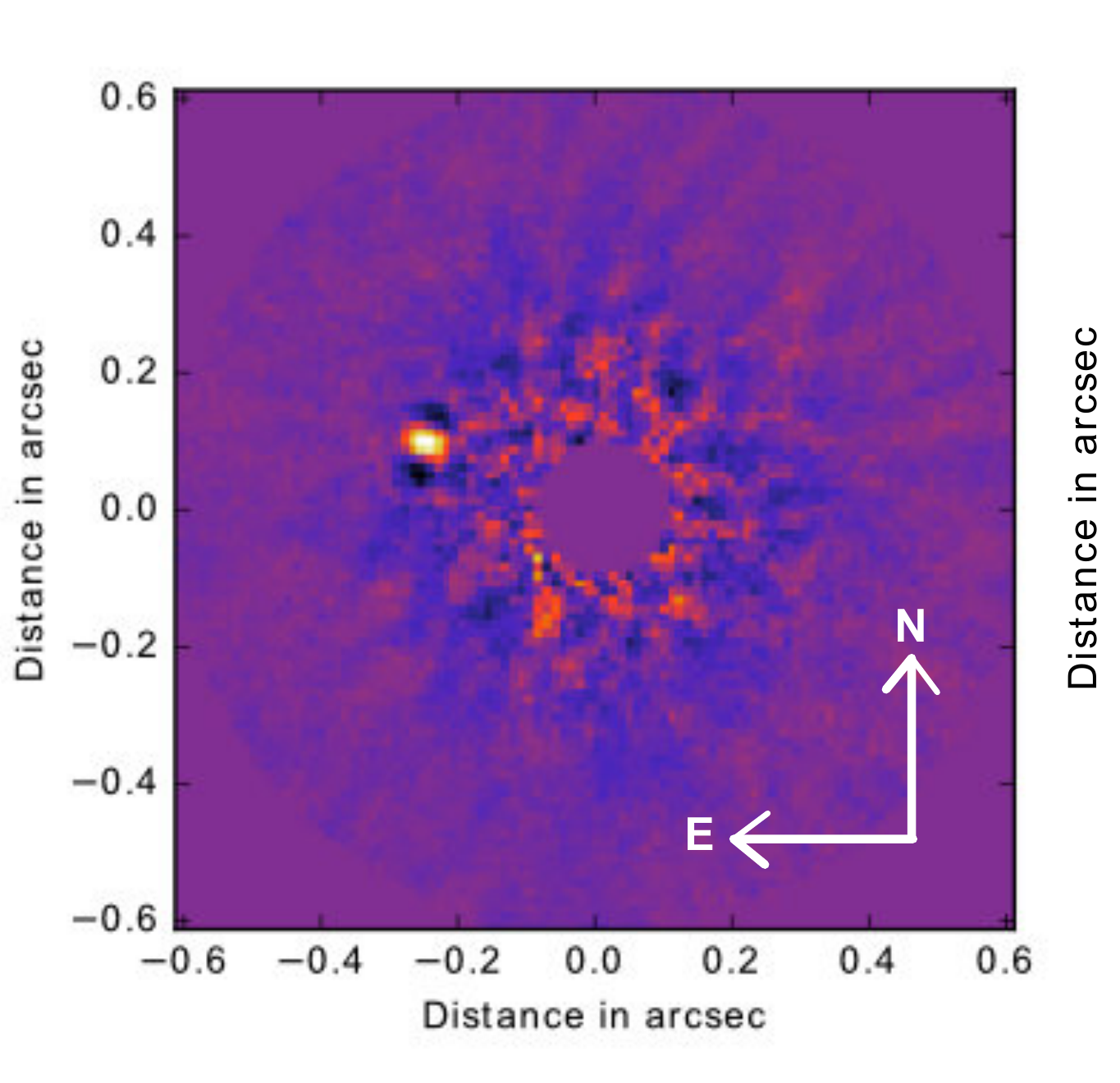}
        \caption{SPHERE H-band coronagraphic image reduced with ADI + PCA, showing the detection of the companion HD\,206893\,B at 270 mas with a S/N of 14.}
        \label{fig_IRDIS}
\end{figure}

The SPHERE High Angular Resolution Debris Disc Survey (SHARDDS) is a high resolution imaging survey aimed at resolving and characterising new debris discs never detected in scattered light \cite[PI: Milli, 096.C-0388, 097.C-0394, see also][]{Wahhaj2016}. This programme is a search for discs around stars within 100\,pc having an infrared excess greater than $10^{-4}$, with the IRDIS subsystem \citep{Dohlen2008} in broad band H and  the apodised Lyot coronagraph of diameter 185mas.  Each target is observed in pupil-stabilised mode to allow angular differential imaging \cite[ADI,][]{Marois2006}. On 5 October 2015, we observed the star HD\,206893. Over the 40\,min effective on-source integration time, we obtained $50^\circ$ field rotation. The atmospheric conditions were average with a mean seeing of 0.9" and a coherence time of 2.8ms, resulting in a Strehl ratio of 85\% in the H band. 
The raw frames were sky-subtracted, flat-fielded and bad-pixel corrected using the SPHERE Data Reduction and Handling pipeline \citep{Pavlov2008}, resulting in a temporal cube of 576 frames with individual integration time 4s. The frames were thereafter re-centred using the four satellite spots imprinted in the image during the centring sequence obtained before and after the 576 frames. With broad-band filters, these satellite spots are elongated. We fitted a 2D Gaussian to each spot and evaluated the star location as the intersection of the two lines joining the centres of opposite satellite spots, as explained in \citet{Wertz2016}, which yields an absolute centring accuracy of 0.2px or 2.5mas. The individual frames of the cube were not re-centred relative to one another because an active centring using the SPHERE differential tip-tilt sensor is dealing with this to an accuracy smaller than what can be obtained from an individual frame-to-frame recentring \citep{Wertz2016}. 
We reduced the images using the principal component analysis algorithm \cite[PCA,][]{Soummer2012}, as implemented in the Vortex Image Processing pipeline (VIP\footnote{ available at \href{https://github.com/vortex-exoplanet/VIP}{https://github.com/vortex-exoplanet/VIP}}, \citealt{Gomez2017}, Fig.~\ref{fig_IRDIS}). In this algorithm, the only free parameter is the number of modes removed. We detect a point source with an S/N of 14 (Fig.~\ref{fig_IRDIS}) at a projected separation of 270.4$\pm 2.6$ mas or 10.4~au. 
\subsection{NICMOS}
\label{sec_data_red_nicmos}

HD\,206893 was observed on 12 June 2007 with the NICMOS instrument on the Hubble Space Telescope (HST). We re-analysed the data (see Appendix \ref{app_nicmos}) and do not detect any point source.

\subsection{NaCo}

HD\,206893 was observed with VLT/NaCo on 8 August 2016, taking advantage of its AGPM coronagraph \cite[e.g.][]{Mawet2005, Mawet2013}. This observation was part of programme 095.C-0937(B) (PI: O. Absil). HD\,206893 was observed in pupil-stabilised mode using the 27.1~mas/pixel plate scale and the \Lp-band filter (3.8\micron). The seeing (0.7-0.8\arcsec) and the coherence time ($\sim$5 ms) were stable throughout the observation. A total of 90 science data cubes of 0.3s (DIT) exposure and 200 (NDIT) frames were obtained, corresponding to 1h30 on-source and a $107^\circ$ total parallactic angle variation. Sky data cubes were obtained every 10-12min. The star was carefully re-centred behind the coronagraph after each sky observation to a $\sim$ 0.2-0.3 pixel accuracy. Four data cubes were also obtained with a shorter exposure time and the star offset from the coronagraph centre (but still behind the AGPM substrate), to obtain unsaturated PSF images. These data cubes were used for photometric calibration and to generate fake companions.\\
After standard calibrations (sky subtraction, flat-fielding and bad pixel correction), the frames were re-centred by fitting a negative gaussian to the AGPM central hole as done in \citet{Absil2013}. The frame selection process, essential to reach the best contrast, kept the 12879 most correlated frames and with the lowest level of residual speckle noise out of the 18000 original frames. These 12879 frames were binned four by four, to yield a final ADI cube of 3219 frames. The final cube was reduced using algorithms based on the PCA implemented in VIP, as shown in Fig. \ref{fig_NaCo} with seven principal components removed, which was found to optimise the companion S/N (approximately six).

\section{Analysis}
\label{sec_analysis}

\subsection{Detection of the companion HD 206893 B}

The astrometry and photometry of the point source detected in the SPHERE and NaCo data set are described in Table \ref{tab_prop}. The ADI algorithms affect the astrometry and photometry of a detected companion. A robust way to estimate them is to use the negative fake companion algorithm \cite[hereafter NEGFC; e.g.][]{Lagrange2010}. For both SPHERE and NaCo, we used NEGFC as implemented in VIP \citep{Wertz2016} with the exploration of the three parameters (radial separation, $PA$ and contrast) performed with a simplex algorithm minimising the residual standard deviation in an aperture at the location of the companion.  
The uncertainties given in Table \ref{tab_prop} combine that on the instrument (plate-scale, north alignement, filter transmission), the centring uncertainty and the measurement uncertainty due to the presence of speckle noise, as detailed in \citet{Wertz2016} for SPHERE and \citet{Absil2013} for NaCo. We used a true north offset of $-1.75\pm0.08^\circ$ for SPHERE \citep{Maire2016} and re-calibrated the NaCo true north against that of SPHERE using the common astrometric field 47 Tuc observed in September 2016 which yielded a value of $0.58\pm0.10^\circ$. The measurement uncertainty was computed by injecting fake companions at the separation of the point source and various azimuths, and by retrieving their astrometry and photometry with the NEGFC algorithm. It dominates the NaCo error budget. For SPHERE, the budget of error is dominated by the 2.5mas conservatively attributed to the centring accuracy. The contrast curves obtained from SPHERE, NaCo and NICMOS are shown in Fig.~\ref{contrast}. They take into account the penalty term coming from small-sample statistics at small separations \citep{Mawet2014}. 
 
This target is at high galactic latitude ($-44^\circ$) and therefore a background contamination within 0.3\arcsec of the star is unlikely. To confirm the object is truly bound, we computed its expected position if it was a background object with no significant proper motion for the date corresponding to the NICMOS and NaCo data. We find a separation and PA of ($0.995\arcsec \pm 0.006$\arcsec; $85.3^\circ \pm 0.58^\circ$) for the NICMOS epoch in 2007 and ($0.174\arcsec \pm 0.003\arcsec$ ; $61.0^\circ \pm 0.6^\circ$) for the NaCo epoch in 2016 also shown in Fig.~\ref{fig_NaCo} right, using a star proper motion of 94.2 mas/yr at PA $89.9^\circ$. 
In the NICMOS data, no point source is detected at high confidence level at the position where the candidate would have been in 2007 assuming it is a background star. Fig. ~\ref{contrast} shows the 5$\sigma$ radial detection limit measured on the combined image. We also repeated the same processing steps described in Section \ref{sec_data_red_nicmos} after injecting a synthetic NICMOS PSF in the raw data at the background star position and at the $3.6\times10^{-5}$ contrast measured on our candidate in the SPHERE data. The injected point source is detected at 7$\sigma$ as shown in Fig. \ref{fig_NICMOS}, demonstrating that the candidate found with SPHERE would have been detected at a high confidence level, if it were a background object. 
In addition, the star was also observed in two other exoplanet surveys: the International Deep Planet Survey \cite[IDPS,][]{Galicher2016} with the Gemini North/NIRI instrument and the Gemini Planet-finding campaign with the Subaru/NICI instrument \citep{Wahhaj2013}. No detection is reported as their discovery spaces start from 0.3\arcsec and 0.5\arcsec respectively (Fig. \ref{contrast}) but they would have been sensitive to a background object with the same magnitude at 1\arcsec.

In the NaCo data, the position of the companion is clearly not compatible within error bars with a background object (Fig.~\ref{fig_NaCo} right). We interpret the changes in projected separation and position angle between the IRDIS and NaCo data as due to the orbital motion of the companion (see section \ref{sec_orbital_motion}).  We can thus confidently assert that this object is bound to HD\,206893.

\subsection{Companion physical properties}

\begin{figure}
\centering
\includegraphics[width=0.9\hsize]{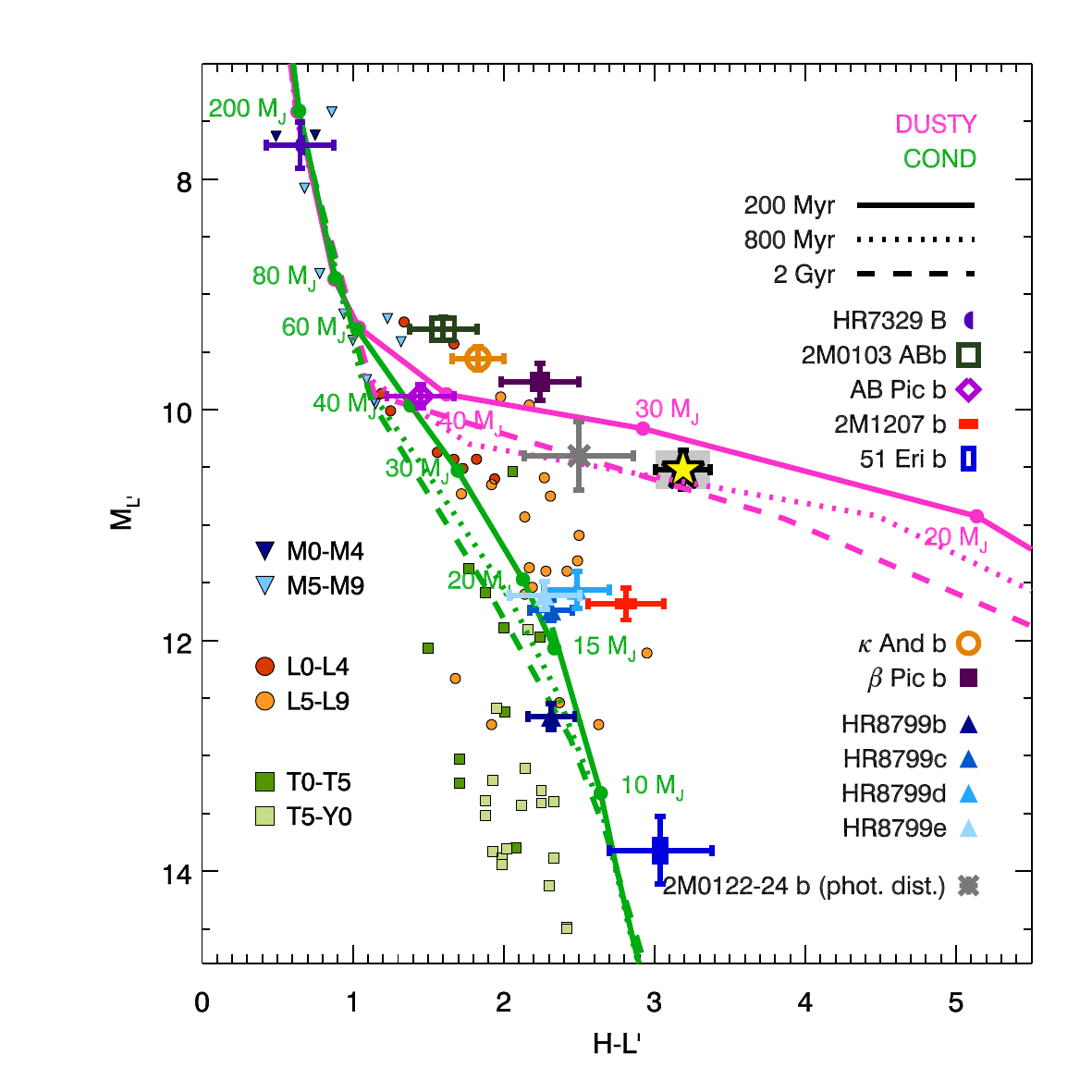} 
\caption 
{ \label{cmd}
Colour-magnitude diagram, obtained from \citet{Galicher2014} with the new photometry of HD\,206893\,B (yellow star), that of 51 Eri \citep{Macintosh2015} and 2MASS\,0122-2439\,B \citep{Bowler2013}. The lines show the isochrones for different ages and evolution models.}
\end{figure}

The companion HD\,206893\,B has a very red colour, with $3.19^{+0.18}_{-0.16}$ mag difference between the H and \Lp{} band. Fig. ~\ref{cmd} compares its position in a colour-magnitude diagram to that of other young companions and field dwarfs \citep{Legett2010,Legett2013}. As the age is debated, we overplotted three isochrones using LYON evolutionary tracks \citep{Chabrier2000, Baraffe2003} for 200 Myr, 800Myr and 2 Gyr. HD\,206893\,B lies among the L5-L9 field dwarf objects, with a similar \Lp magnitude as 2MASS\,0122-2439\,B  \citep{Bowler2013} but a redder colour. This makes it the reddest object among young and dusty L dwarfs in the field, which is likely due to a dusty atmosphere although interstellar or disc reddening cannot be ruled out. Using an age of 200\,Myr (respectively 800\,Myr, 2\,Gyr) and the AMES-Cond models \citep{Baraffe2003}, the H-band contrast of HD\,206893\,B implies an object of 24\Mjup{} and effective temperature 1230K (50\Mjup with 1330K, and 75\Mjup{} with 1420K respectively). 

\subsection{Relation to the debris disc and orbital motion}
\label{sec_orbital_motion}

\begin{figure}
        \centering
  \includegraphics[width=\hsize]{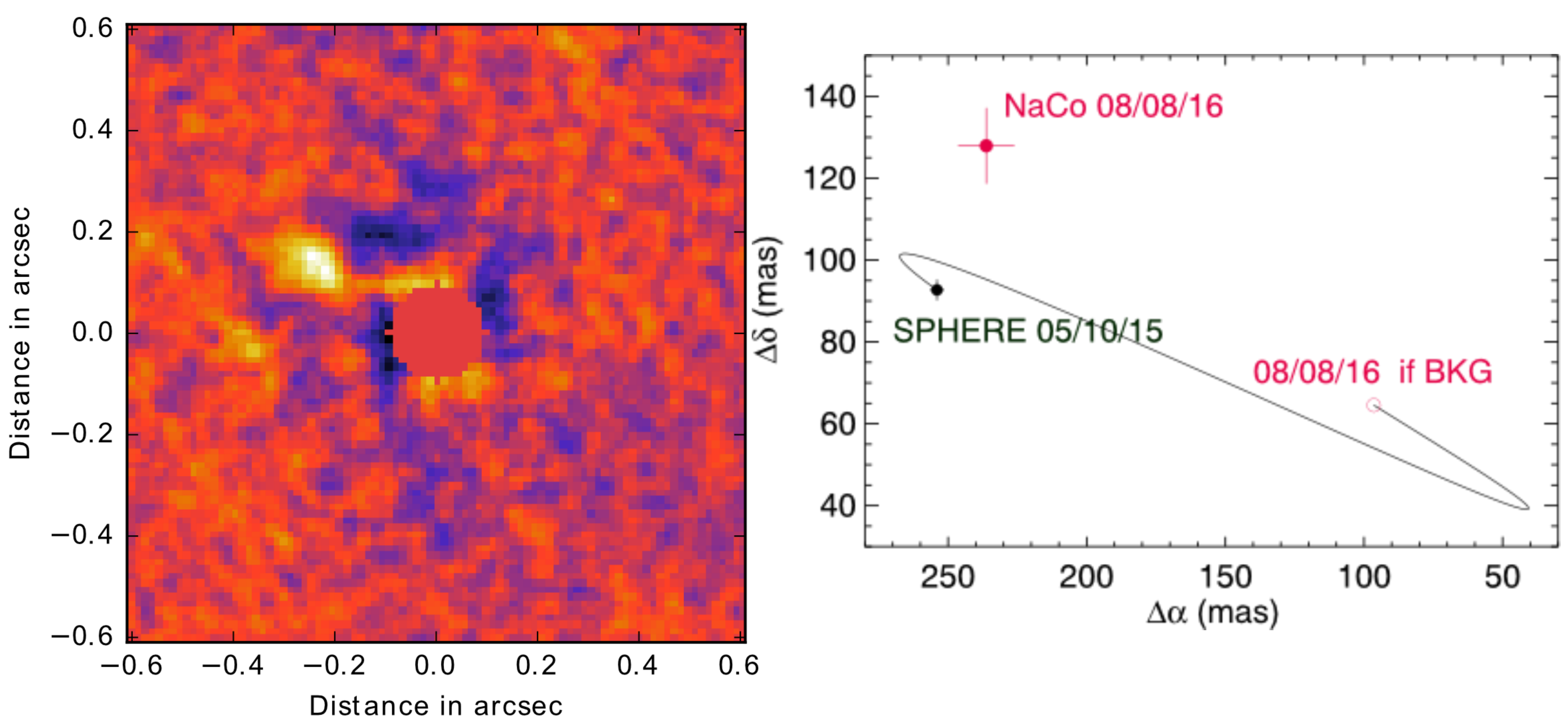}
        \caption{Left: Image of NaCo reduced with PCA-ADI after removing seven principal components, which was found to optimise the companion S/N ($\sim 6$). Right: common proper motion analysis. The black line displays the motion of the companion if it was a background source. The uncertainty is given at $1\sigma$.}
        \label{fig_NaCo}
\end{figure}

\begin{figure}
\centering
\includegraphics[width=0.9\hsize]{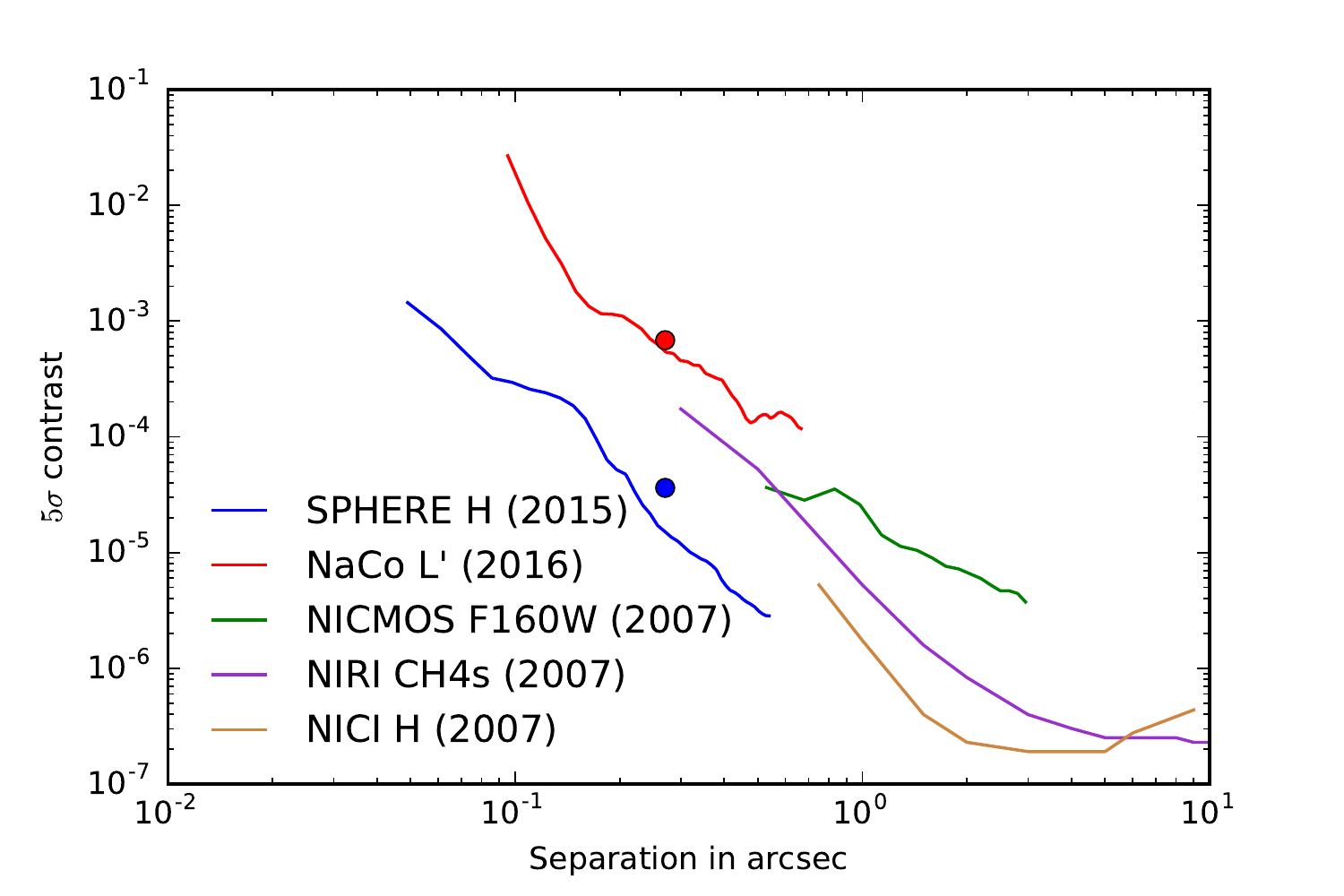}
\caption{ Constrast curves at $5\sigma$. The red and blue dots show the positions of the companion in the NaCo and SPHERE data respectively. }
\label{contrast}
\end{figure}

The debris disc is marginally resolved with Herschel/PACS with an inclination of $40^\circ\pm10^\circ$ along the PA  $60^\circ \pm10^\circ$. This is presented in Appendix \ref{app_disc} along with a modelling concluding on a disc inner radius of 50\,au. With a projected separation of $10.4\pm0.1$\,au, the companion appears therefore to be interior to the disc, with a PA consistent with an orbit in the same plane as the disc.
As shown in Fig. \ref{fig_NaCo} right, the companion moved during the 306 days separating the SPHERE and NaCo detections. We applied the methods laid in \citet{Pearce2015} to constrain the orbit of a companion imaged over short orbital arcs. With the two epochs, the linear sky motion is $0.05 \pm 0.01$\arcsec/year at a position angle of $-27{^\circ}^{+13}_{-17}$.  We derived thereafter the parameter $B$ and $\phi$ of \citet{Pearce2015}, that proves useful to constrain the possible orbits. Including astrometric, distance, and 10\% stellar mass uncertainties we find $B=0.33^{+0.19}_{-0.14}$ and $\phi=90^\circ \pm 15^\circ$. The constraints are weak as the NaCo astrometry is relatively poor; in particular we cannot constrain the ascending node to test whether the orbit is indeed coplanar with the disc. If we assume a low eccentricity orbit, the semi-major axis is about 12\,au (37yr period). Conversely, for a high eccentricity, if the apocentre distance is restricted to be below 50au (i.e. the approximate disc inner edge), this implies that the companion is currently at apocentre, that is with a semi-major axis below 10\,au. If we restrict the companion to lie within a projected separation of 500\,mas in June 2007 to avoid detection by NICMOS (see Fig. \ref{contrast}), the marginalised constraints on specific parameters do not improve much.
If the companion truncates the disc, then the semi-major axis is expected to be $\sim$1/3 of the inner edge radius \citep{Holman1999}, so 15-20~au (0.4-0.5\arcsec). Thus, while there is no evidence to suggest that the companion's orbit is unusual, more precise astrometry is required to constrain the orbital elements and draw conclusions about the relation between the companion's orbit and the disc. 

This architecture is strikingly similar to the HR\,2562 system \citep{Konopacky2016}, where a $30\pm15$ \Mjup{} brown dward was detected at 20\,au of its host star, within a debris disc of inner radius $\sim75$\,au.

\begin{table}
\caption{Properties of the system HD\,206893}             
\label{tab_prop}      
\centering                     
\begin{tabular}{p{2.5cm} >{\centering}p{2.5cm} >{\centering}p{3.3cm} >{\centering}p{1.cm} }        
\hline\hline                 
Property & HD\,206893 &  HD\,206893\,B & Ref. \tabularnewline
\hline                        
Distance (pc) &  \multicolumn{2}{c}{$38.34 \pm 0.8$} & 1 \tabularnewline
 \multirow{2}{2.5cm}{Proper motion (mas/yr)} &  \multicolumn{2}{c}{$\mu_\alpha \times \text{cos }\delta =93.67 \pm 0.66$} & 1 \tabularnewline
                                                            &  \multicolumn{2}{c}{$\mu_\delta =0.33 \pm 0.37$} & 1 \tabularnewline
Age (Gyr) &  \multicolumn{2}{c}{0.2-2.1} & 2 \tabularnewline
Spectral type & F5V &  L5-L9 dwarf & 1 / 3,4\tabularnewline
H mag & 5.69 & $16.79 \pm 0.06$ & 1 / 3 \tabularnewline
\Lp mag & 5.52 & $13.43^{+0.17}_{-0.15}$ & 1 / 4 \tabularnewline
Mass & $1.24 M_{\odot}$ &  24\tablefootmark{a} / 50\tablefootmark{b} / 73\tablefootmark{c}   \Mjup & 5 / 3 \tabularnewline
$T_\text{eff}$ (K) & 6486 &  1200\tablefootmark{a} / 1310\tablefootmark{b} / 1380  \tablefootmark{c} & 2 / 3 \tabularnewline
 \multirow{2}{*}{Separation (mas)} & & $270.4 \pm 2.6$ & 3 \tabularnewline
                                                    & & $268.8  \pm 10.4$ & 4 \tabularnewline 
 \multirow{2}{*}{PA ($^\circ$) } & & $69.95 \pm0.55$  & 3 \tabularnewline 
                                                & &  $61.6 \pm 1.9$ & 4 \tabularnewline 
\hline                                   
\end{tabular}
\tablefoot{For an age of
\tablefoottext{a}{0.2 Gyr,}
\tablefoottext{b}{0.8 Gyr, and}
\tablefoottext{c}{2 Gyr respectively.}
}
\tablebib{
(1)~\citet{VanLeeuwen2007}; (2)~\citet{Zuckerman2004,Holmberg2009,Pace2013,David2015}; (3)~ This work: VLT/SPHERE (05/10/15);
(4)~This work:VLT/NaCo (08/08/16); (5)~\citet{David2015};
}
\end{table}

\section{Conclusions}
\label{sec_conclusion}

This letter presents the detection of a low-mass companion orbiting at a projected separation of 10 au around the F5V star HD\,206893 as part of the SHARDDS survey, thanks to VLT/SPHERE  high contrast capabilities. The object was confirmed by VLT/NACO and proven not to be a background source by HST/NICMOS. With an H-band contrast of 11 mag, evolutionary models suggest the object could be a 25\Mjup{} brown dwarf if the system is 200 Myr, or twice as massive for a 800Myr system. Along with its \Lp contrast of 7.9, the object appears very red, and the closest to L5-L9 field dwarfs in a colour-magnitude diagram. The orbital motion is detected and suggests an orbital period of $\sim37$ yr in case of low eccentricty. 

In addition, we report the detection of the disc, through its thermal emission with Herschel/PACS at a position angle of $\sim60^\circ$ almost aligned with the projected position of the companion. This system is therefore reminiscent of the cases of HR\,8799, HD\,95086 or $\beta$ Pictoris where one of several GPs have been detected in orbit inside a Kuiper belt analog. It is the second brown dwarf detected in the inner hole of a debris disc after HR\,2562.

Several aspects make this system very attractive for future characterisation. The contrast is well within range of current extreme AO instruments, enabling spectral identification. The orbital motion is fast enough to allow orbit monitoring which can bring constraints on the dynamical mass of the object. Deeper observations may detect the scattered light of the disc and confirm the faint emission seen in our image, to understand if the companion is responsible for the inner truncation of the disc at about 50 au, or possibly reveal asymetries and clumps resulting from interactions between the disc and the brown dwarf or possible yet undiscovered planets.

\begin{acknowledgements}
JM is supported by the ESO fellowship programme. He thanks R. Galicher for his re-reduction of the NIRI data, and O. Wertz for his help with VIP and NEGFC. EC is supported by NASA through Hubble Fellowship grant HST-HF2-51355 and HST-AR-12652  awarded by STScI, operated by the AURA, Inc., for NASA under contract NAS5-26555. OA is a F.R.S.-FNRS Research Associate. The research leading to these results was partly funded by the European Research Council under the European Union's Seventh Framework Programme (ERC Grant Agreement n.\ 337569), and by the French Community of Belgium through an ARC grant for Concerted Research Action. GMK is supported by the Royal Society as a Royal Society University Research Fellow.   MCW and LM are supported by the European Union through ERC grant 279973. VC acknowledges J. Smoker and M. Espinoza for their help with NaCo. VC is supported by the Millennium Science Initiative (Chilean Ministry of Economy) through grant RC130007. CdB acknowledges support from the Mexican CONACyT research grant CB-2012-183007.
\end{acknowledgements} 

\vspace{-1cm}
\bibliography{biblio}   
\Online
\begin{appendix} 

\section{NICMOS non detection}
\label{app_nicmos}

\begin{figure}
        \centering
  \includegraphics[width=\hsize]{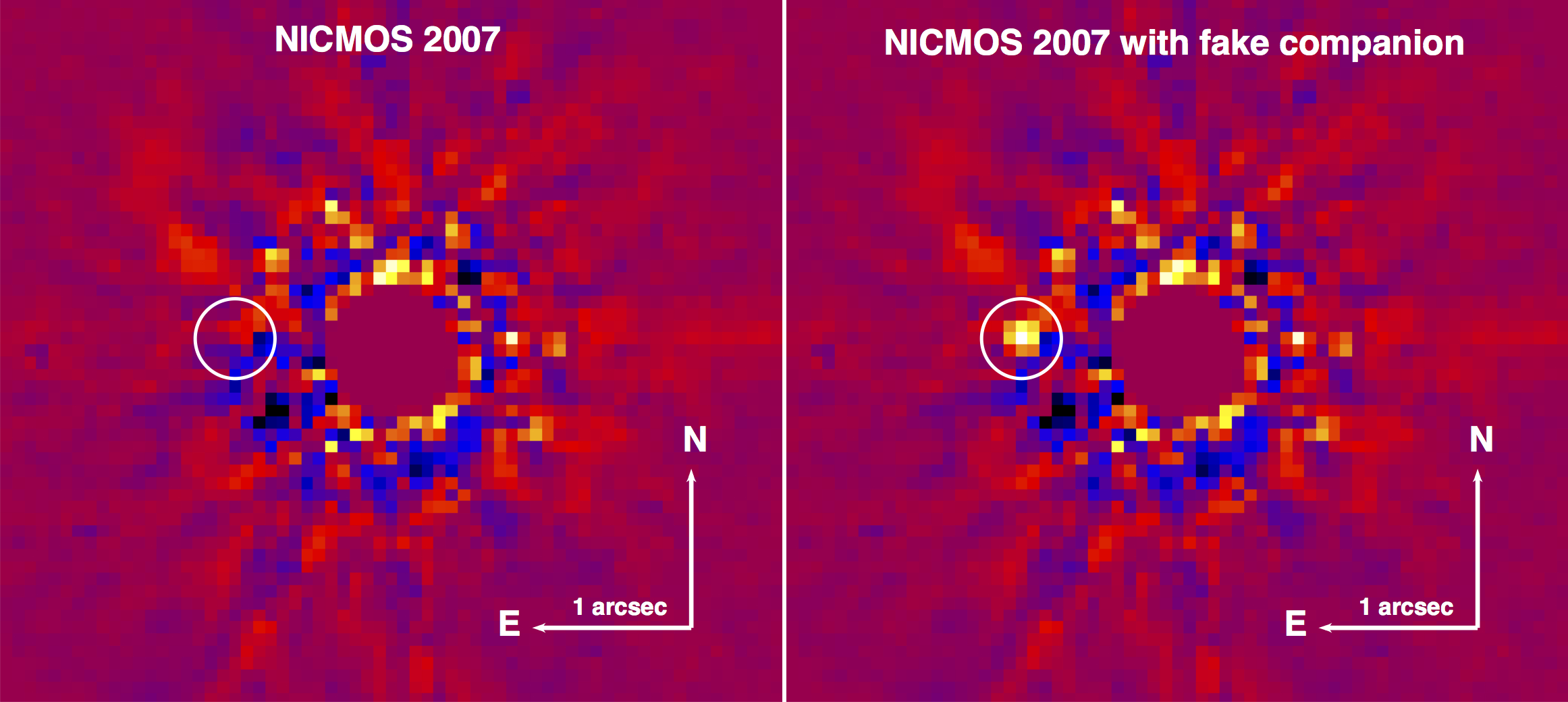}
        \caption{F160W NICMOS 2007 coronagraphic images, processed with RDI + PCA without injecting a fake companion (left) and after injection of a fake companion of contrast 11.1mag at the location where the candidate would have been in 2007 assuming it is a background object with no proper motion (white circle).}
        \label{fig_NICMOS}
\end{figure}

HD\,206893 was observed on 12 June 2007 with the NICMOS instrument on the \emph{Hubble Space Telescope} (HST), as part of a survey looking for debris discs around nearby stars (PI: J.Rhee, GO-11157). The data were obtained with the mid-resolution NIC2 channel (plate scale 0.07565\arcsec/px) with a coronagraph of radius 0.3\arcsec, in two filters F160W and F110W. The target was observed at two orientations of the telescope separated by $\sim30^\circ$, to enable PSF subtraction with roll differential imaging. We reprocessed the F160W archival dataset (centre wavelength 1.6006\micron, FWHM 0.4012\micron) with the same PSF subtraction method as used in the Archival Legacy Investigations for Circumstellar Environment (ALICE) programme \citep{Soummer2014,Choquet2016}. 
We used the KLIP algorithm \citep{Soummer2012} on PSF libraries composed of images from multiple reference stars. After bad pixel correction, we selected the 454 images (from 78 reference stars) the most correlated with each of HD\,206893's exposures, out of a reference star library assembled with the ALICE pipeline. This selection favoured images from 78 different stars chosen mostly from the two dominant HST programmes in the initial library (programmes 11157 and 10176). Fig. ~\ref{fig_NICMOS} (left) shows the combined image after subtracting synthetic PSFs computed from the 55 strongest eigenmodes of the library. No point source is detected at the position where the candidate would have been in 2007 if it were a background object without proper motion. Our detection limits (Fig \ref{contrast} and \ref{fig_NICMOS} right show that the point source would have been detected in that case. This re-analysis shows that the point source is co-moving with the star.

\section{The disc around HD 206893}
\label{app_disc}

\begin{figure}
        \centering
        \includegraphics[width=0.9\hsize]{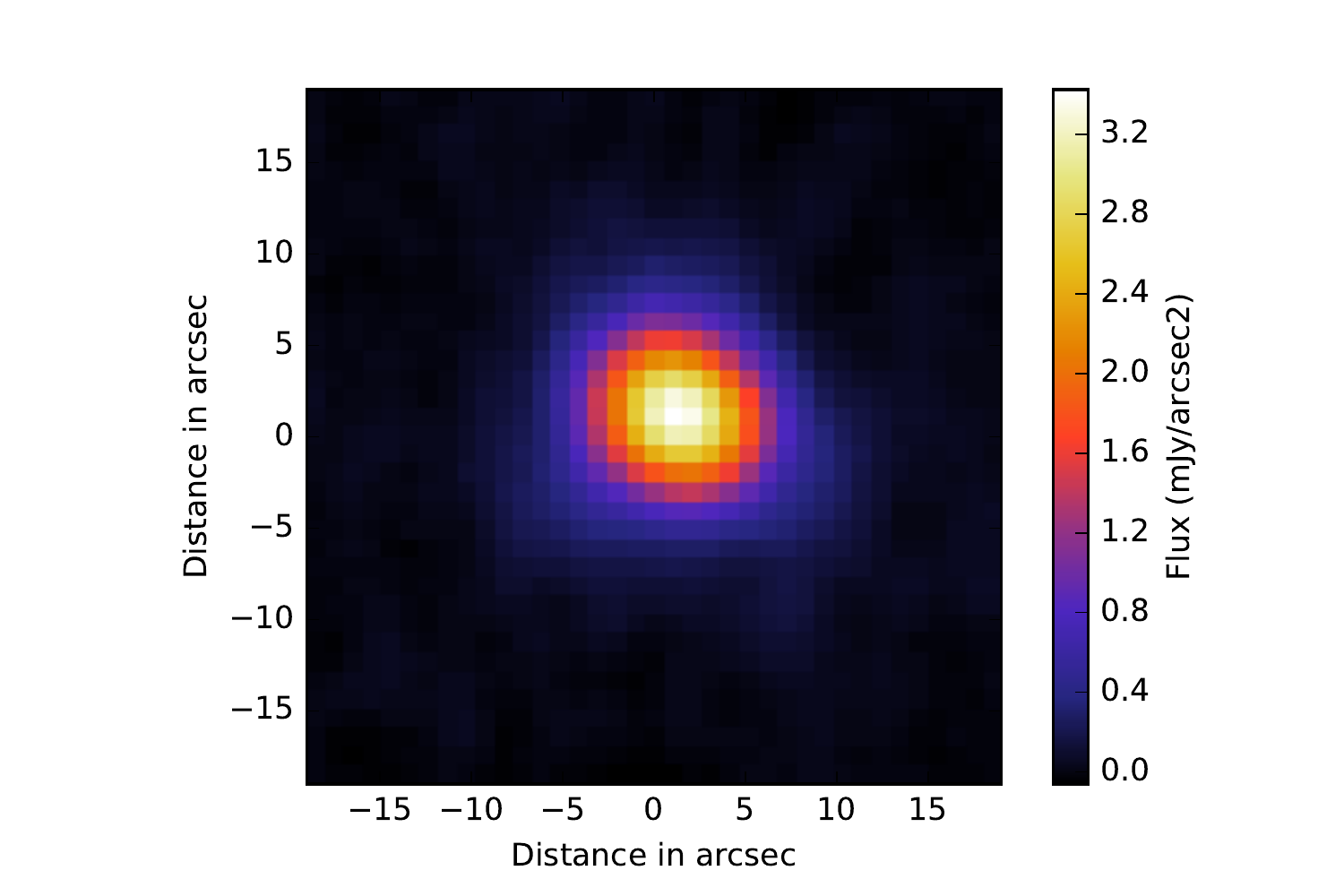}
        \includegraphics[width=0.9\hsize]{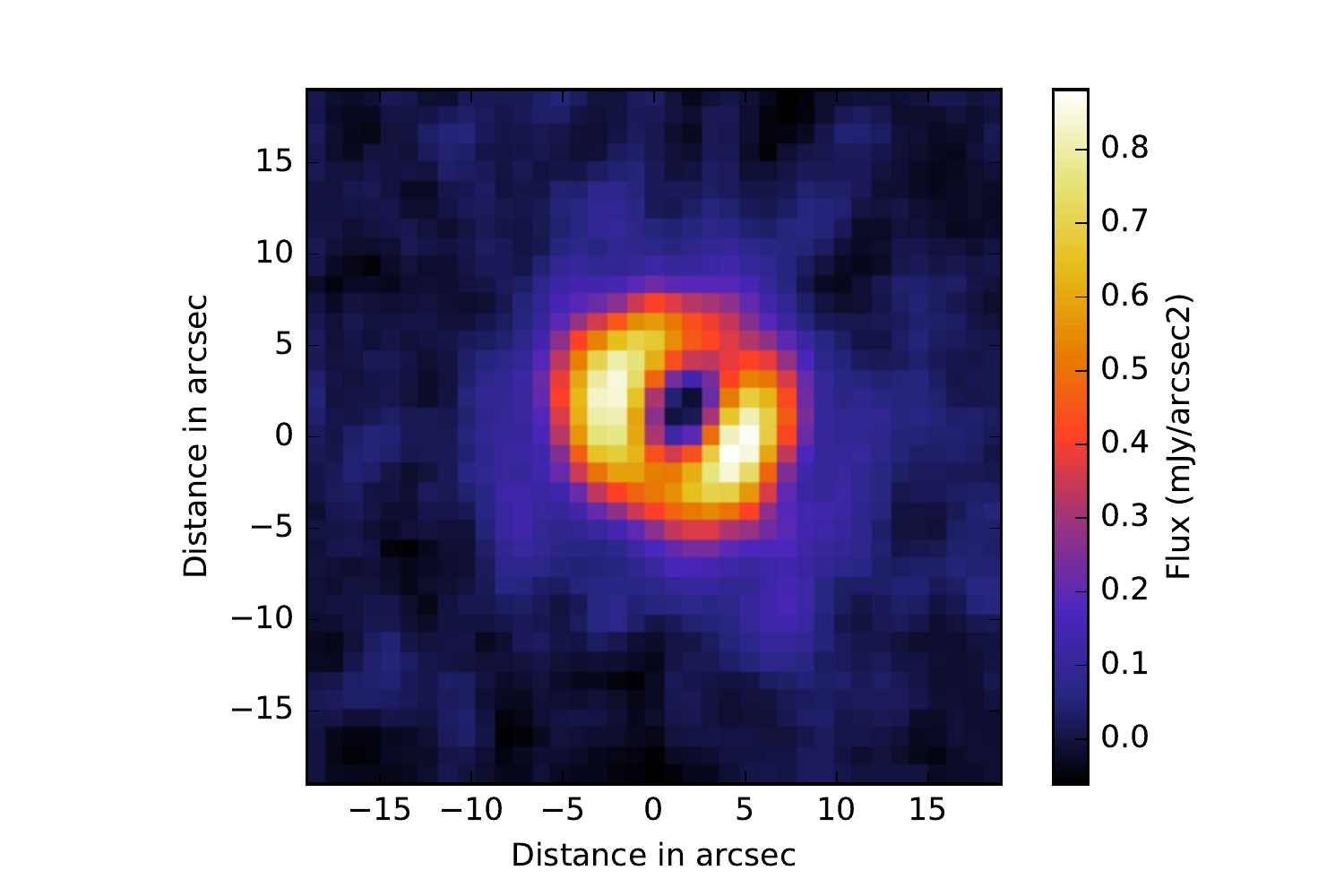}
        \caption{Herschel / PACS image of the star HD\,206893 at 70\micron{} (top). The bottom image shows the residuals after subtraction of the Herschel PSF, showing the source is not point-like.}
        \label{fig_disk_herschel}
\end{figure}

Archival Herschel PACS data were obtained and modelled in the same way as \citet{Kennedy2012_99Her,Kennedy2012_coplanar_disk}. The 70\micron{} image is shown in Fig. \ref{fig_disk_herschel} (top image). The subtraction of a scaled calibration observation (PSF) clearly shows that the source HD\,206893 is not point-like (bottom image) and reveals the approximate disc extent and position angle. The debris disc is only a few beams across so constraints on the disc properties are relatively poor. Our disc best fitting model assumes a temperature $T_{\rm disk}=288 r^{-0.4}$, with $r$ in au, and has a decreasing power-law surface density distribution of $\Sigma \propto r^{-0.5}$ extending from 50 to 200\,au. This is compatible with previous modelling by \citet{Moor2011} who proposed a modified blackbody model for the disc with a temperature of 49K and a radius of 49 au. This is also compatible with the colder (48K) dust population in the double component disc model of \citet{Chen2014}. The disc is inclined (from face-on) by about 40$^\circ$ at a position angle of $\sim60^\circ$ (East of North). The uncertainties on these angles are on the order of 10$^\circ$

\begin{figure}
        \centering
        \includegraphics[width=\hsize]{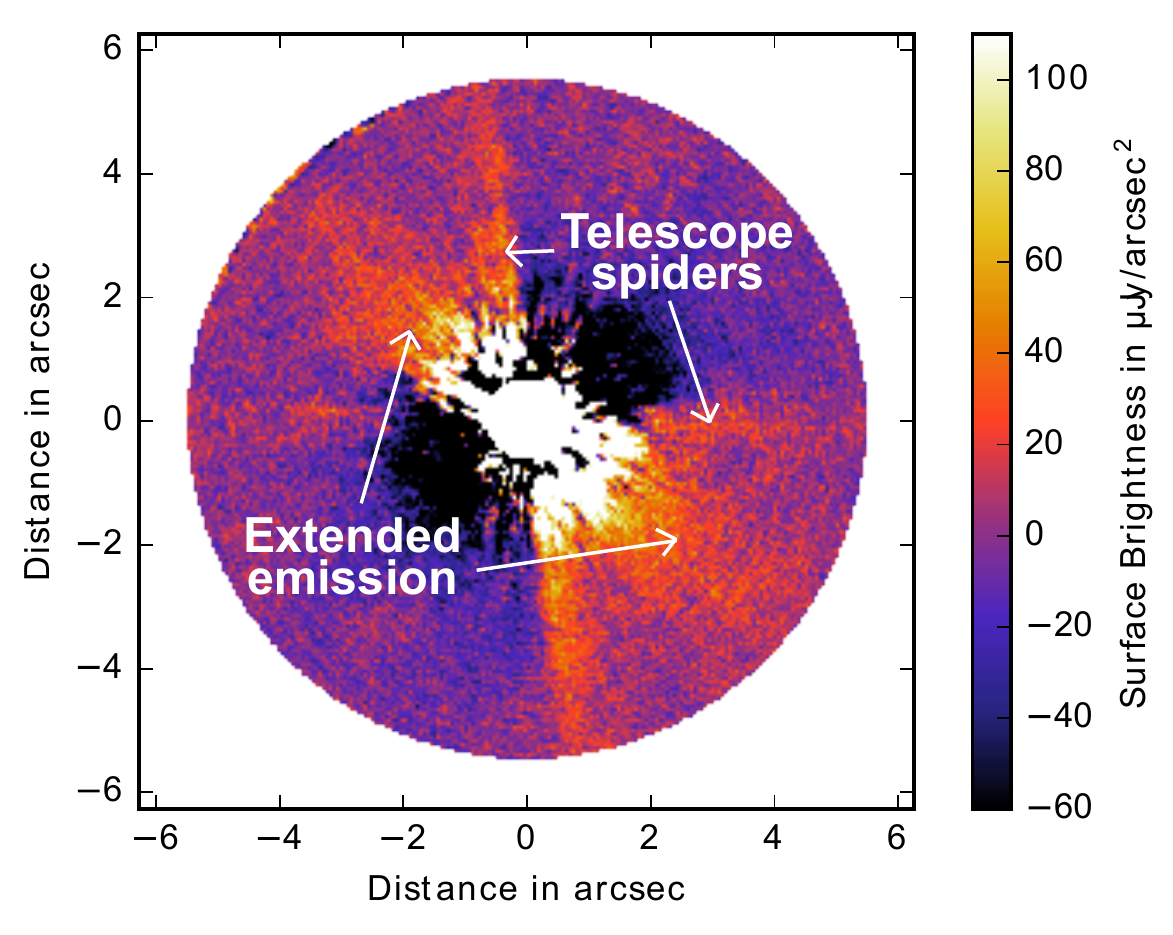}
        \caption{SPHERE H-band image in its complete $12\arcsec\times12\arcsec$ field of view after a classical ADI reduction, in a linear colour scale in ${\mu}$Jy/arcsec$^2$. It shows a faint and extended emission along the same PA as the Herschel residual image likely coming from the disc.}
        \label{fig_disk_IRDIS}
\end{figure}

To try to reveal the disc in scattered light, we reduced the SPHERE images using classical ADI (Fig.~\ref{fig_disk_IRDIS} showing the whole $12\arcsec\times12\arcsec$ field of view of SPHERE/IRDIS) to maximise the sensitivity in the backround and limit flux losses induced with more aggresive reductions \citep{Milli2012} and binned the pixels by a factor two. We see a faint extended emission along the PA $\sim60^\circ$ with a surface brighntess of $\sim0.05 $mJy/arcsec$^2$. This faint emission is detected from $\sim1.5\arcsec$ (60 au) up to $\sim$4-5\arcsec (150-190\,au) where the background noise starts to dominate.We could confidently rule out spurious emission along that PA coming from the diffraction pattern from the spiders or the elongation of the PSF due to the wind at the ground level or at higher altitudes. Furthermore the PA is compatible with the Herschel/PACS residual image. We therefore tentatively attribute this signal to the scattered light of the debris disc with a S/N of approximately one. 

 \end{appendix}

\end{document}